\pdfoutput=1

\documentclass[11pt]{article}

\usepackage{acl}

\usepackage{times}
\usepackage{latexsym}

\usepackage[T1]{fontenc}

\usepackage[utf8]{inputenc}
\usepackage{booktabs}
\usepackage{array}
\usepackage{multirow}
\usepackage{xcolor}
\usepackage{adjustbox}
\usepackage{graphicx}
\usepackage{fixltx2e}
\usepackage{hyperref}
\usepackage[inline]{enumitem}
\usepackage{multicol}

\newcommand{\header}[1]{\vspace*{1.75mm}\noindent\textbf{#1}.}
\newcommand{\subheader}[1]{\vspace {0.75mm}\indent\textbf{#1}.}

\author{Oleg Litvinov$^1$, Ivan Sekuli\'c$^2$, Mohammad Aliannejadi$^1$, Fabio Crestani$^2$ \\
  $^1$University of Amsterdam, The Netherlands \\
  $^2$Università della Svizzera italiana, Switzerland}

\title{Analyzing Coherency in Facet-based Clarification Prompt Generation for Search}

\begin{document}

\maketitle
\begin{abstract}
Clarifying user's information needs is an essential component of modern search systems.
While most of the approaches for constructing clarifying prompts rely on query facets, the impact of the quality of the facets is relatively unexplored.
In this work, we concentrate on facet quality through the notion of \emph{facet coherency} and assess its importance for overall usefulness for clarification in search.
We find that existing evaluation procedures do not account for facet coherency, as evident by the poor correlation of coherency with automated metrics.
Moreover, we propose a coherency classifier and assess the prevalence of incoherent facets in a well-established dataset on clarification.
Our findings can serve as motivation for future work on the topic.
\end{abstract}

\vspace{-4mm}
\section{Introduction}
\begin{figure}
    \centering
    \vspace{-2mm}
    \includegraphics[width=\columnwidth]{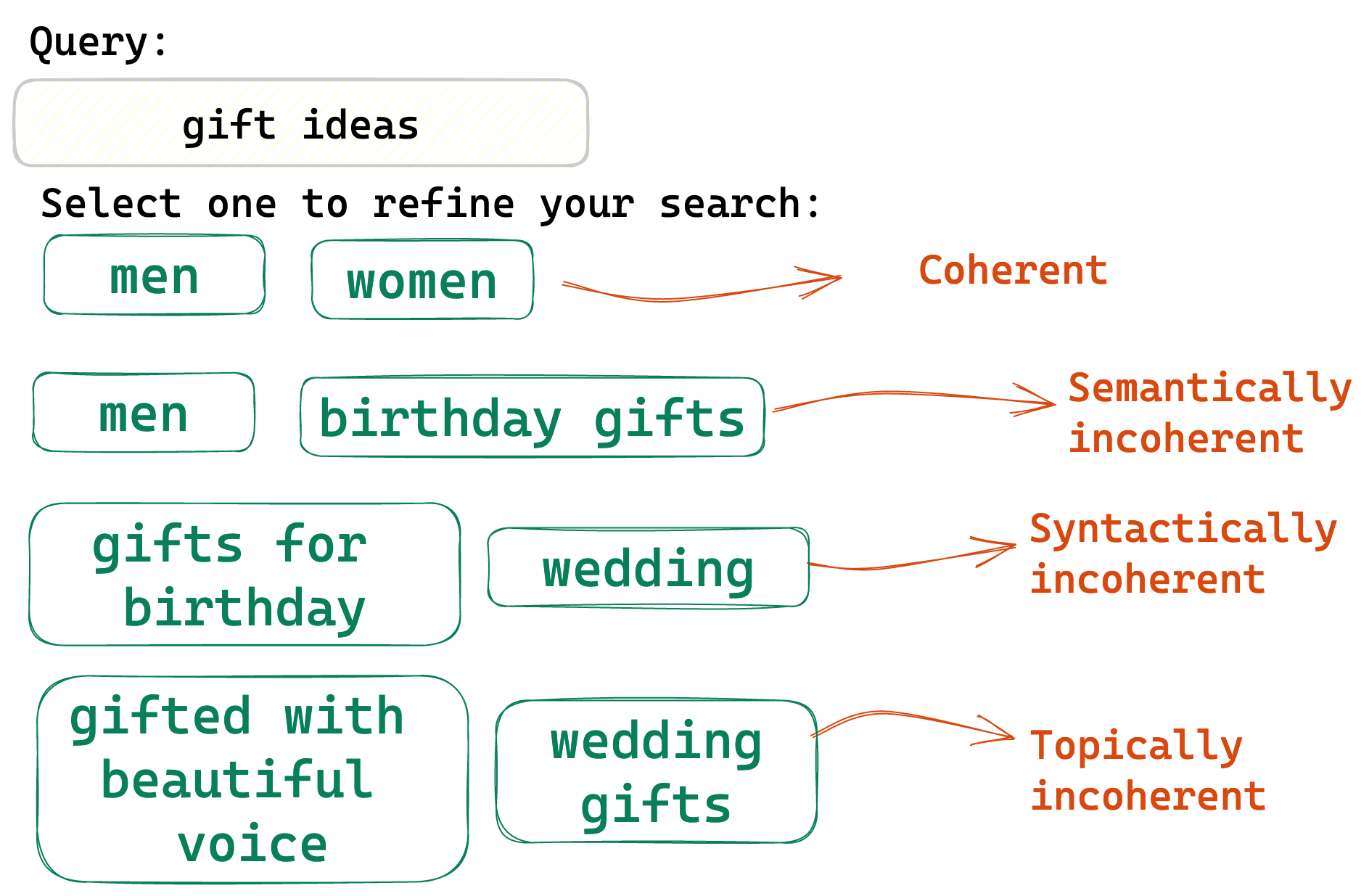}
    \vspace{-4mm}
    \caption{Examples of coherent and incoherent facets for clarification in search.}
    \label{fig:createit2}
    \vspace{-5mm}
\end{figure}

\vspace{-2mm}
Clarification has emerged as an important component of modern search systems.
By elucidating the underlying user information need, the search system can retrieve more relevant information and provide the user with a more satisfactory overall search experience.
Recently, clarification in search has been studied in the framework of traditional search engines in the form of clarification panes~\cite{zamani2020mimics} and under the conversational search paradigm in the form of clarifying questions~\cite{aliannejadi2019asking,sekulic2021towards}.
While clarification prompts can take different forms depending on the search medium, they are generally composed of a clarifying question with one or more query facets as options for resolving potential ambiguity, as shown in Figure~\ref{fig:createit2}.
Recent studies assess the quality of clarification panes from different aspects such as diversity and coverage of query facets~\cite{tavakoli2022mimics}; however, they do not consider the inter-dependency of facets, nor their relationship with the question. In this work, we argue that the query facets of a question depend on each other and the question.
We analyze several aspects of facet extraction methods in depth, with a focus on \emph{coherency}, where we hypothesize that more coherent query facets would lead to higher quality questions and improved user experience. 

The query facets that serve as the basis for clarifying prompt construction have previously been extracted from various sources, including mining large-scale query logs~\cite{zamani2020generating}, constructing clarifying questions based on subtopics identified by crowdsourcing workers~\cite{aliannejadi2019asking}, and extracting them from a list of documents retrieved in response to the original user query~\cite{hashemi2021learning}.
However, although all these methods show promising performances, the quality and the comparison of various facet-extraction methods for clarification in search remains an unresolved research question.\looseness=-1

To this end, we provide an in-depth analysis of facets extracted from query logs and facets generated by a large language model (LLM) based on the documents retrieved in response to the original user query.
First, we define a notion of facet \emph{coherency}, which indicates whether the facets we base our clarification prompt on correspond to a similar level of abstraction and topical coherency.
For example, as indicated in Figure \ref{fig:createit2}, given the user's query ``gift ideas'', facets \emph{men} and \emph{women} would be deemed as coherent.
However, facets \emph{women} and \emph{birthday gifts} are incoherent, even though relevant to the topic, while \emph{birthday gifts} and \emph{wedding gifts} are coherent.


Moreover, we challenge the established metrics for measuring the performance of generated facets, such as BLEU and BERTScore, by pointing out that a low score on those NLG metrics does not necessarily mean the facet extraction method is faulty.
Finally, we introduce the notion of \emph{coherency} as one of the key components of overall facet quality.
We discuss how low \emph{coherency} of the facets can impact user satisfaction with the search system and
 define several types of incoherent facets, as depicted in Figure \ref{fig:createit2}, including syntactical, semantical, and logical incoherency.
Moreover, we train a coherency classifier aimed at distinguishing coherent and incoherent facet sets.


We are interested in addressing one research question, \textbf{RQ1}: \textit{How does facet coherency impact the overall quality of clarifying questions? And how can we predict it?}
In answering \textbf{RQ1}, our contributions can be summarized as follows: \looseness=-1
\begin{itemize}[nosep,leftmargin=*]
    \item We introduce a notion of \emph{facet coherency} and analyze different facet extraction methods, challenging the established performance metrics.
    \item We manually annotate a dataset of \emph{facet coherency} and \emph{facet quality}, analyzing the problem in depth. \looseness=-1
    \item We build a classifier to assess facet coherency and identify challenges in the existing work.
\end{itemize}

\vspace{-2mm}
\section{Methodology}
\vspace{-2mm}

\header{Facet Extraction}
\label{sec:method:facets}
Given a query $q$ and retrieved list of documents $D = [d_1,\dots,d_i,\dots,d_N]$, where $d_i$ is $i$-th document in a list and $N$ is the maximum number of documents considered, the task is to extract a set of facets $F$.
The set $F$ contains several facets, consisting of single-term or multi-turn facets, and represents potential facets of $q$.
Formally, $F = \{f_1,\dots,f_j,\dots,f_M\}$, where $M$ is the number of facets extracted.
We note that, while the facet set has no particular order of facets, we depict $f_j$ as one of the facets in the set.
Moreover, $f_j = [f_j^1,\dots,f_j^l,\dots,f_j^L]$, where $f_j^l$ depicts $l$-th term of the facet $f_j$ and $L$ is the number of terms in the facet.
Finally, we define the task of facet extraction as:
    $F = \theta(q, D)$,
    where $\theta$ is a mapping form ($q$, $D$) to the facet set $F$.
    
In this paper, we model $\theta$ with a generative LLM.
Specifically, inspired by \citet{hashemi2021learning}, we utilize BART~\cite{lewis2019bart} --- a seq2seq LLM model for text generation.
The input of the model consists of $q$ and $D$, separated by a special token, and followed by the language modeling target $F$. 
We fine-tune the \texttt{BartForConditionalGeneration}~\cite{hf} model on ($q$, $D$, $F$) triplets.
Following \cite{hashemi2021learning}, the number of facets in a set $M$ is known in advance. 
Thus, we stop the facet generation procedure after $M$ facets via intentional generation.
While the performance of the facet generation method is not central to this study, as we aim to rather model facet coherency, we present the results in terms of automated NLG metrics in Section~\ref{sec:results:facet_gen}.

\header{Dataset}
\label{sec:method:dataset}
To study facet extraction methods, we base our analysis on the aforementioned MIMICS~\cite{zamani2020mimics} dataset. 
Each query $q$ is associated with a clarification pane, consisting of a clarifying question and a set of clickable query facets $F$.
As we train our facet extraction method on this facet set, we refer to it as ground truth facet set $F_{GT}$ in the following sections.
Furthermore, the collection contains the top 10 documents retrieved in response to each query, which we utilize as our document list $D$.
Specifically, we base our experiments on MIMICS-Click, the largest of the datasets in the MIMICS collection, containing 414,000 triplets of ($q$, $D$, $F_{GT}$).


\header{Facet Coherency}
Facet quality can be assessed from different points of view~\cite{hashemi2021learning,kong2016precision}, including term overlap with the reference facet sets, user engagement, impact on retrieval, and lexical soundness.
Arguably, the most important measure is user satisfaction.
While user satisfaction is often related to satisfying the user's information need by retrieving a relevant list of documents in response to a user's query, we, in this work, look at user satisfaction through the lens of coherency.
Specifically, we argue that for a complete user experience, the interaction between the user and the system needs to be natural and not confusing.
For example, Figure~\ref{fig:createit2} depicts several possible facet sets.
We argue that, while certain facet sets might still be useful for clarification from the retrieval perspective, users would benefit from coherent sets of facets, rather than potentially confusing, unrelated sets.

We extend the notion of coordinate facets given by \citet{kong2013extracting} to define \emph{coherency}.
Coherent facets are on the axis of subtopics, i.e., semantically aligned.
An example given in Figure~\ref{fig:createit2} shows facets ``men'' and ``birthday gifts'' as an example of semantically incoherent pair, whereas their semantically coherent pairs might be ``men'' and ``women'' or ``birthday gifts'' and ``wedding gifts''.
This definition aligns with \citet{kong2013extracting}.
Moreover, we define syntactical incoherency as facets in a facet set having significant differences in phrasing, i.e., appended parts of a query vs. raw facet, as indicated in Figure~\ref{fig:createit2}.
Finally, we have topical incoherency, i.e., unlike in semantic incoherency where a set of facets is simply not on the same axis, the facets extracted are dealing with completely different word meanings.
An example is shown in Figure~\ref{fig:createit2}, where ``gift'' is for one facet interpreted as a present, and for the other as a special ability to do something. \looseness=-1

Next, we define our procedure for acquiring weakly-supervised data for training facet coherency classifiers and our crowdsourcing-based evaluation of facet coherency and quality.
We define facet set as of high quality if it is likely to improve the search experience, focused on resolving the ambiguity of the original query.




\subheader{Expert Point-wise Annotation} 
\label{sec:method:expert}
In order to both gain further insight into facet coherency and to acquire data for training coherency classifiers, we analyze in-depth facet sets from MIMICS~\cite{zamani2020mimics}.
To this end, we manually inspect each of the possible clarifying questions in MIMICS, together with the associated facets.
We label as coherent the facet sets that align with the above definition of coherency.
Other facet sets, including facets that contained logical mistakes, are deemed as incoherent.
In addition, we accept small typos such as missed whitespace or one letter added or missed (except 
 for abbreviations).
 This annotation procedure results in 217 facet sets labeled for coherency.
 

We curate a weakly annotated set of more than $1200$ samples with coherency labels, 
identifying a number of clarifying questions in MIMICS, which are consistently associated with well-structured coherent facet sets.
We sampled up to 50 examples for each question and annotated each ground truth and generated values in the same way as described above. 
If coherency is stated for more than 95\% of examples, all of the facets associated with the question are deemed coherent.
Moreover, facet sets containing at least one same facet twice or facets containing the original query are labeled as incoherent. \looseness=-1

\subheader{Crowdsourcing-based Pair-wise Annotation}
 In order to study the impact of facet coherency on clarification in search, we assess facet quality and facet coherency in a pair-wise setting. 
Specifically, annotators were asked to assess which facet set is better (in terms of the definitions given below): one generated by LLM ($F_{BART}$) or the ground truth from the MIMICS dataset ($F_{GT}$).
Our intentions with this experiment are threefold:
\begin{enumerate*}
    \item We aim to compare LLM-based facet generation ($F_{BART}$ to query log-based facet extraction methods ($F_{GT}$);
    \item We aim to assess the correlation between facet coherency and facet quality;
    \item We aim to assess the correlation between facet coherency and performance in terms of NLG metrics.
\end{enumerate*}

We utilize Amazon Mechanical Turk\footnote{\url{https://mturk.com}} as our annotation platform, with several steps to ensure higher quality annotations: workers need at least 1000 HITs and a 95\% approval rate; workers perform well on the curated test set; workers are based in The US, to mitigate language-barrier challenges for understanding the task.
The workers were provided appropriate compensation, resulting in well above the minimum wage in The US.
We shuffle the order of the dataset to reduce positional bias.
In total, crowdsourcing-based annotation resulted in $199$ comparisons.


\header{Coherency Classifier}
\label{sec:method_binary}
Here, we describe our novel classifier for predicting the coherency of a set of facets $F$.
Formally, the classifier aims to model function $\gamma$: $s = \gamma(F)$, where $s \in [0, 1]$.
The facet set is deemed coherent if $s > 0.5$, and incoherent otherwise.
We utilize the classifier to extrapolate knowledge acquired through the annotation procedure described in Section~\ref{sec:method:expert} on the MIMICS dataset as a whole.
Thus, we fine-tune the classifier on the same 3k samples of facet coherency.
We employ BERT~\cite{devlin2018bert} model from \texttt{bert-base-cased}, HuggingFace~\cite{hf} as our classifier $\gamma$.
We fine-tune the model for 4 epochs and set the early stopping for the validation loss with the patience hyperparameter of 2.
The model is fine-tuned on the aforementioned data with a stratified 70/15/15 split for train, validation, and test set, respectively.
The effectiveness of the classifier and assessed prevalence on the entire MIMICS dataset is presented in Section~\ref{sec:results:classifier}.\looseness=-1



\vspace{-2mm}
\section{Results and Discussion}
\vspace{-2mm}


\header{Facet Extraction}
\label{sec:results:facet_gen}
\begin{table}[]
\centering
\caption{Results for the query facet generation and coherency evaluation experiments. \label{tbl:facet}}
\vspace{-2mm}
\adjustbox{max width=\columnwidth}{%

 \begin{tabular}{ccccccccc}
 \toprule
  & \multicolumn{4}{c}{Set BLEU} & BERTScore & METEOR & \multicolumn{2}{c}{Coherency score} \\ 
 \midrule
 M & 1-gram & 2-gram & 3-gram & 4-gram & & & GT & BART \\
 \midrule
2 & 0.63 & 0.57 & 0.54 & 0.51 & 0.91 & 0.47 & 0.39 & 0.31 \\ 
 3 & 0.54 & 0.47 & 0.43	& 0.40 & 0.89 & 0.37 & 0.25 & 0.30 \\
 4 & 0.50 & 0.41 & 0.37 & 0.34 & 0.89 & 0.33 & 0.16 & 0.33 \\
 5 & 0.45 & 0.33 & 0.29 & 0.26 & 0.87 & 0.24 & 0.07 & 0.27 \\
 
 \bottomrule
 \end{tabular}

}
\vspace{-2mm}
\end{table}
Table~\ref{tbl:facet} shows the results of our facet generation method described in Section~\ref{sec:method:facets}, compared to the ground truth facet sets in MIMICS.
We make several observations from the results.
First, as measured by all of the NLG metrics, i.e., $n$-gram BLEU, BERTScore, and METEOR,  the performance of our facet extraction method drops with the increase of the number of facets $M$.
This phenomenon is rather expected, as the higher number of facets opens more room for mistakes.
Similarly, with the higher $n$-gram count in the BLEU score, the performance drops.
Similar tendencies were reported by \citet{hashemi2021learning}.
The observed results indicate that LLM-based facet generation while showcasing decent overall performance, performs poorly with the higher number of terms it needs to generate, as measured by the aforementioned NLG metrics.



Regarding coherency prediction (the last two columns of Table~\ref{tbl:facet}), we observe several trends.
First, the coherency score for ground truth facets taken from MIMICS, i.e., facets extracted from query logs by internal Bing algorithms \cite{zamani2020mimics}, degrades with the increase of the number of facets.
This suggests that the higher the number of facets, the harder it is to maintain the coherency of said facets.
Second, the drop in the coherency score is 
lower for the generated facets, suggesting that the BART model generates somewhat more stable sets of facets.
This could be because language models generate the most probable sequences, based on specific decoding strategies, potentially causing results of higher coherency.

\header{Facet Coherency}
\begin{table}
\centering
\caption{Crowdsourcing results for pair-wise comparison of ground-truth and generated facets. The $\dagger$ indicates a statistically significant difference (trinomial test, $p$-value $< 0.01$). \label{table:annotation}}
\vspace{-2mm}
\adjustbox{max width=\columnwidth}{%
 \begin{tabular}{llll}
 \toprule
  Experiment & Ground-truth wins & Tie & Generated wins \\ 
 \midrule
 Facet coherency & 29\% & 43\% & 28\% \\
 Facet quality $\dagger$ & 60\% & 24\% & 16\% \\ 
 \bottomrule
 \end{tabular}
 }
 \vspace{-4mm}
\end{table}
Table \ref{table:annotation} presents the annotation results collected from 2 workers for each HIT. 
If both workers voted for the same set of facets as more coherent (or of higher quality in a separate study), we count that as a win for the method, the set of facets is generated with.
If two workers had opposing views on which set of facets is more coherent (of higher quality), we count that as a tie.
We perform trinomial test~\cite{bian2011trinomial}, an alternative to the Sign and binomial tests that take into account ties, to test for statistical significance.
With a null hypothesis of equal performance, we observe no statistically significant difference for facet coherency between $F_{GT}$ and $F_{BART}$, thus accepting the null hypothesis in terms of coherency.
In terms of facet quality, we reject the null hypothesis, i.e., $F_{GT}$ are deemed of significantly higher quality than generated facets $F_{BART}$.
As we define facet quality as their ability to help in search, i.e., by resolving potential ambiguity and leading the user toward the desired answer, these findings suggest that facets from MIMICS help more in search than those generated by BART.
As the facets in MIMICS are constructed by Bing's internal algorithms, and based on large-scale real-world query logs, this finding is somewhat expected. \looseness=-1



\header{Coherency Prediction}
\label{sec:results:classifier}
We evaluate the coherency prediction model on 15\% of the annotated data as described in Section~\ref{sec:method_binary}.
Classifier on generated sets $F_{BART}$ reaches an accuracy score of 0.75 and F1-macro of 0.74. The same setup on MIMICS ground truth $F_{GT}$ has an accuracy of 0.82 and an F1-macro of 0.79.
All considered classification metrics are slightly better for $F_{GT}$, probably, due to a larger dataset collected. 
Finally, we employ our coherency classifier to assess the prevalence of incoherent facet sets in MIMICS, as well as generated facets for all of the MIMICS queries.
The procedure suggests that $F_{GT}$ are incoherent in 78\% and $F_{BART}$ in 93\% of the cases.

\vspace{-2mm}
\section{Conclusions}
\vspace{-2mm}
In this paper, we analyzed the concept of facet \emph{coherency} for clarification in search. 
We identified several types of incoherent facet sets and confirmed the evaluation gap when existing NLG metrics-based approaches are used.
As indicated by qualitative analysis, we find that various types of incoherency could potentially affect user experience and model performance at different levels.
In the future, we aim to extend our work to assess the direct impact of facet quality on user satisfaction in search, through a user study.

\section*{Limitations}
In this paper, we study the significance of coherency in the quality of generated facets for query clarification.
Our study is limited to the publicly available dataset, requiring more extensive analysis of multiple datasets on the topic. We note that certain biases inherited from the training data can influence the outcome of our model, as well as the trained coherency classifier. Further study on the potential biases and approaches to mitigating them is necessary in the future. Moreover, we note the need for further investigation of other classifiers, as well as models relying on the knowledge of large language models for the coherency prediction task. 
Finally, we acknowledge the fast-paced development of LLM and note the potential improvements of coherency classifiers by utilizing the latest developments in the open-source LLM field.

%
%
\bibliographystyle{plainnat}
\bibliography{mybib}

\appendix

\section{Correlation of Coherency and NLG Metrics}
We find that traditional NLG metrics, such as BLEU, METEOR, and BERTScore do not necessarily reflect the quality and coherency of the extracted facets.
We look into a subset of the annotated data where the facets generated are highly different from the ground truth facets and the subset where they are similar.
Moreover, we filter out all the non-coherent facets to assess whether the NLG metrics are misleading.
Naturally, BERTScore is significantly higher in the subset where facets are similar to the ground truth ($p$-value $< 0.01$).
However, since both of the subsets contain only coherent facet sets, we conclude that a facet set can be coherent and of high quality, but yield low NLG metrics.

Table \ref{table:correlation_ex} shows several cases where generated facets were scored poorly by NLG metrics, although they are both of high quality and coherent. 
For example, for the query ``1982 mustang'' Set BLEU, BERTScore, and METEOR are low while both ground truth and generated facets are coherent. 
For the ``new call of duty game,'' NLG metrics are low while the facets generated are coherent. 
Contrariwise, the "police sales" query shows high NLG metrics while the facets generated are incoherent --- the ``school bus sales'' facet is not related to the query.
Thus, we conclude that careful evaluation of query facets is needed, as scores given by NLG metrics might be misleading and result in biased performances.

\begin{table*}[]
\caption{Examples of contradicting metrics.}
\label{table:correlation_ex}
\resizebox{\textwidth}{!}{%
\begin{tabular}{p{2cm}p{2.7cm}p{2.7cm}p{1.3cm}p{1.3cm}lp{1.5cm}p{1.5cm}}
\toprule
Query & $F_{GT}$ & $F_{BART}$ & BLEU 1-gram & BERT Score & METEOR & Coherent ($F_{GT}$) & Coherent ($F_{BART}$) \\ \midrule
1982 mustang & coupe; hatchback & specs; for sale & 0.36 & 0.82 & 0 & True & True \\ \midrule
new call of duty game & new call of duty zombie game; new call of duty ghost game & pc; ps4 & 0.00 & 0.83 & 0 & False & True \\ \midrule
police sales & police car sales; police motorcycle sales; police boat sales & police car sales; police motorcycle sales; school bus sales & 0.90 & 0.98 & 0.64 & False & False \\ \bottomrule
\end{tabular}%
}
\end{table*}

\end{document}